\documentclass[11pt,a4paper]{article}
\pdfoutput=1

\usepackage[]{Packages}

\RequirePackage{placeins}
\RequirePackage{ragged2e}

\usepackage{Definitions}

\allowdisplaybreaks[1]

\preprint{CERN-PH-TH/2011-292}

\newcommand{\OfficialTitle}{Twisted Masses and Enhanced Symmetries:\\the A\&D Series}

\title{\vspace{2cm}
  {\huge   \textbf{\OfficialTitle}}
}

\hypersetup{pdfauthor={Domenico Orlando and Susanne Reffert},pdftitle={\OfficialTitle}}

\author{
  \begin{minipage}{.8\linewidth}
    \vspace{1cm}
    \begin{center}
      {\small \textbf{Domenico Orlando} and \textbf{Susanne Reffert} }
    \end{center}
    \vspace{1cm}
    \begin{minipage}{\linewidth}\centering
      {\itshape \footnotesize 
        Theory Group, Physics Department, CERN\\CH-1211 Geneva 23, Switzerland
      }
    \end{minipage}
  \end{minipage}
}

\date{}

\begin{document}

\setstretch{1.1}

\numberwithin{equation}{section}

\begin{titlepage}

  \maketitle

  \thispagestyle{empty}

  \vfill

  \abstract{ We study new symmetries between \( A \) and \( D \) type quiver gauge theories with \mbox{different} numbers of colors.
    We realize these gauge theories with twisted masses via a brane construction that reproduces all the parameters of the Gauge/Bethe correspondence.  }
\vfill

\end{titlepage}

\section{Introduction}

The Gauge/Bethe correspondence~\cite{Nekrasov:2009ui, Nekrasov:2009uh, Nekrasov:2009rc} %
identifies the supersymmetric vacua of an $\mathcal{N}=(2,2)$ gauge theory in two dimensions with the spectrum of a superselection sector of a Bethe--solvable integrable spin chain.
In this note we discuss how the symmetry group of the integrable model is reflected in the gauge theory and we argue that it corresponds to a \emph{new global symmetry} that relates quiver gauge theories with different numbers of colors.

\bigskip

\FloatBarrier

\begin{figure}[b]
  \centering
  \begin{tikzpicture}%
    \node[anchor=south west,inner sep=0] (image) at (0,0){\includegraphics[width=8cm,height=5cm]{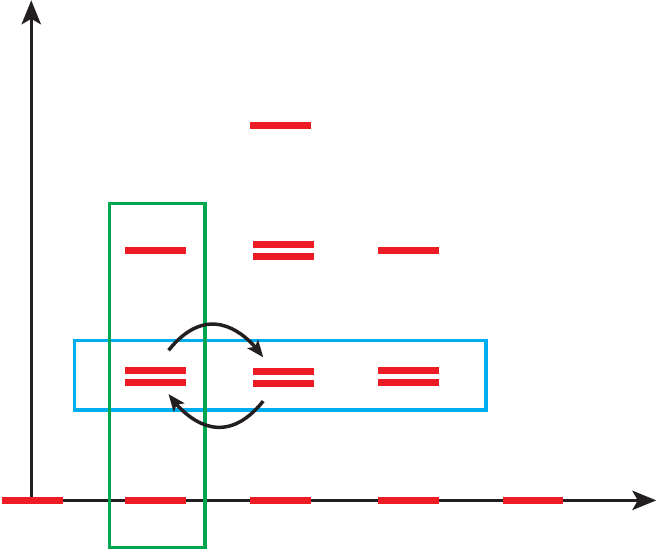}};
    \begin{scope}[x={(image.south east)},y={(image.north west)}]

      \begin{small}
        \node at (0,0.95) {\( E \) }; 
        \node at (0.95,0.05) {\( N \) };
        \foreach\x in {0,1,...,4}{\node at (0.05+\x*.19,0.05){\( \x \)};}
        \node at (.35,.47) {\( S^+ \) };
        \node at (.35,.18) {\( S^- \) };
      \end{small}
        
      \begin{footnotesize}
        \node[anchor=west] at (.75,.32) {%
          \begin{minipage}{5em}
            \emph{\( SU(2) \)\\multiplet}
          \end{minipage}}; 
        \node[anchor=west] at (.17,.7) {%
          \begin{minipage}{5em}
            \emph{gauge\\theory}
          \end{minipage}};
      \end{footnotesize}

    \end{scope}
  \end{tikzpicture}
  \caption{The \( A_1 \) symmetry on the \textsc{xxx}${}_{1/2}$ spin chain for \( L = 4 \). The horizontal arrows show the action of the \( S^\pm \) operators, changing the magnon number \( N \), preserving the energy. The spectrum can be organized into multiplets of \( SU(2) \) (horizontal box) or by magnon number (vertical box).}
  \label{fig:symmetry-spectrum}
\end{figure}
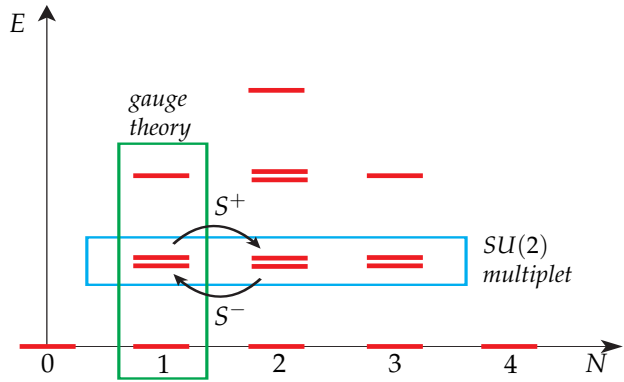

The action of the symmetry group of the integrable system is
exemplified in the sketch in Figure~\ref{fig:symmetry-spectrum}. The
lines represent the energy levels for the eigenstates of an \textsc{xxx} spin
chain of length \( L =4 \) which has symmetry group $SU(2)$, as a function of the number of magnons \(
N \). The \( A_1 \) global operators act between states with different
values of \( N \) preserving the energy (horizontal arrows). This
implies that the full spectrum of the chain is organized into \( SU(2)
\) multiplets (horizontal box). The Gauge/Bethe correspondence
identifies the states with a fixed number of magnons (vertical box)
with the ground states of a gauge theory (in this case two-dimensional
\( \mathcal{N} =2^* \) \( U(N) \) with \( L \) flavors). The \( A_1 \)
action thus translates into a symmetry between gauge
theories with different numbers of colors. The set of ground states of all the \( U(N) \) gauge theories for \( 0
\le N \le L \) generates the tensor product of \( L \) copies of the
fundamental representation of \( SU(2) \)~\cite{Orlando:2010aj},
\begin{equation}
  \set{ \text{ground states}} \simeq V = \bigotimes_{k=1}^L \setC^2 \, .  
\end{equation}

Symmetries in \( V \) (such as between the \( N = 1 \) and \( N = 3 \) states in Figure~\ref{fig:symmetry-spectrum}) indicate \textsc{ir}--equivalent gauge theories whose brane realizations are typically related by Hanany--Witten moves~\cite{Orlando:2010uu}.%
\footnote{Different \textsc{ir} equivalences %
  based on the Gauge/Bethe correspondence have been  discussed in~\cite{Chen:2011sj}.} %
In the following, we will study two-dimensional \( A_r \)~and~\( D_r \)~type quiver gauge theories which correspond to integrable models with symmetry group $SU(r+1)$ and $SO(2r)$. The raising and lowering operators act on the nodes of the quivers by raising or lowering the number of colors of the corresponding gauge group by one.

\bigskip

The \emph{twisted masses} of the gauge fields are a vital ingredient for reproducing the parameters of the integrable system.
Schematically, the twisted mass \( \wt m \) for a field \( X \) appears in the action as:
\begin{equation}
  \int \di^4 \theta \, X^\dagger \eu^{\theta^- \bar \theta^+ \wt m + \text{h.c.}} X \, .
\end{equation}
The twisted masses are important because they induce an effective twisted superpotential in which a discrete set of \emph{normalizable} ground states lives. 

A mechanism to incorporate these twisted masses into a string theory realization was introduced in~\cite{Hellerman:2011mv} for the simplest case corresponding to an \textsc{xxx}$_{1/2}$--spin chain with $SU(2)$ symmetry.
The construction employs a stack of
D2--branes suspended between parallel NS5--branes plus a stack of $L$~D4--branes, similar to~\cite{Hanany:1997vm, Orlando:2010uu}, which are
placed into a so-called \emph{fluxtrap} background which is T--dual to
the Melvin or fluxbrane solution~\cite{Melvin:1963qx}. This background
also serves to turn on an $\Omega$--background where
$\varepsilon_1=-\varepsilon_2\in\mathbb{R}$ and can be easily
generalized to general values of
$\varepsilon_{1,2}$~\cite{Reffert:2011dp}. In this note, we will prove the versatility of this approach by generalizing the mechanism to systems with 
Yangian symmetry $Y_{SU(r+1)}$ (\emph{i.e.} the $A_r$~series) and $Y_{SO(2r)}$ (\emph{i.e.}  the $D_r$~series). While the generalization to
the $A_r$--case is straightforward and consists in placing $(r + 1)$
NS5--branes with stacks of D2--branes between them into the fluxtrap
background (Section~\ref{sec:a-series}), reproducing the $D$~series necessitates the use of a more
exotic object, namely an S--dual of an O5--plane on a D5--brane (Section~\ref{sec:d-series}).

While the symmetry group of the integrable model is not directly manifest in the gauge theory setting, it appears naturally in the string theory construction as an \emph{enhanced symmetry for coincident \NS5--branes}. 
One of the reasons that this extra symmetry of the gauge theory has been overlooked in the past is that without twisted masses, which are a relatively exotic and therefore rarely used ingredient, the normalizable states on which the symmetry acts in a tractable way disappear.

\bigskip 

Before delving into the intricacies of the string theory construction
of the Gauge/Bethe correspondence, it is necessary to remind the reader of the
parameters of the integrable system we are aiming to reproduce (for a
more pedagogical introduction, see~\cite{Orlando:2010uu}). The
simplest case of a spin chain
consists of a 1d lattice of length $L$ with either an up or a down
spin at each position $k$. They generate the
two-dimensional fundamental representation of \( SU(2) \) and, with a
proper choice of boundary conditions, the Hamiltonian commutes with
the action of a global \( SU(2) \). As a result, the full spectrum of
the chain can be organized in terms of representations of the symmetry
group. In this note, we want to move on to more general systems, in
particular those with $A_r$ and $D_r$ symmetry as opposed to $A_1$. A
spin chain with a symmetry group of rank $r$ will have $r$ different
particle species. For each of them, the chain has an effective length
$L_a$, $a=1,\dots, r$. $N_a$ denotes the number of particles of
species $a$ in a given superselection sector. We can furthermore
choose a different representation of the symmetry group at each
position $k$ in the chain. $\Lambda_k^a$ denotes the highest weight
of the representation at position $k$ for the $a$-th species. On top
of that, one can also turn on \emph{inhomogeneities} at each position of the
chain; these are parametrized by $\nu_k^a$. The eigenvectors of the
spin chain are labelled by the \emph{rapidities} $\lambda_i^{(a)}$, which satisfy the Bethe
Ansatz equation~\cite{Reshetikhin:1987xx}:
\begin{equation}
  \label{eq:general-NBAE}
  \prod_{k=1}^{L_a} \frac{ \lambda_i^{(a)} + \frac{\imath}{2} \left(  \Lambda_k^{a} + \nu_k^a \right) }{ \lambda_i^{(a)} - \frac{\imath}{2} \left( \Lambda_k^a - \nu_k^a \right)} = \prod_{\substack{(b,j)=(1,1)\\(b,j)\neq (a,i)}}^{(r,N_b)} \frac{ \lambda^{(a)}_i - \lambda^{(b)}_j + \frac{\imath}{2} C^{ab}}{\lambda^{(a)}_i - \lambda^{(b)}_j - \frac{\imath}{2} C^{ab}} \, , \hspace{1.5em} a=1,2, \dots, r \, ; \hspace{.7em} i= 1,2, \dots, N_a \, ,
\end{equation}
where $C^{ab}$ are the elements of the Cartan matrix of the symmetry group. %
The observation of~\cite{Nekrasov:2009ui, Nekrasov:2009uh, Nekrasov:2009rc} is that the same equations describe the ground states of a corresponding two-dimensional quiver gauge system on a circle (for details see Table~\ref{tab:dictionary}).  
It is important to note that the twisted masses (which are the counterparts of the parameters \( \Lambda, \nu, C \)) induce an \emph{effective twisted superpotential} $\widetilde W$ whose minima correspond to the normalizable ground states. These ground states are in one-to-one correspondence with the spectrum of the chain and will be identified in the following with \textsc{bps} configurations in the string theory construction (see~\cite{Hellerman:2011mv}).

\FloatBarrier

\section{The $A$~series}
\label{sec:a-series}

Generalizing the $SU(2)$ or $A_1$ case to $A_r$ is very
straightforward. We will therefore use this section to remind the
reader of the fluxtrap construction~\cite{Hellerman:2011mv}.
 
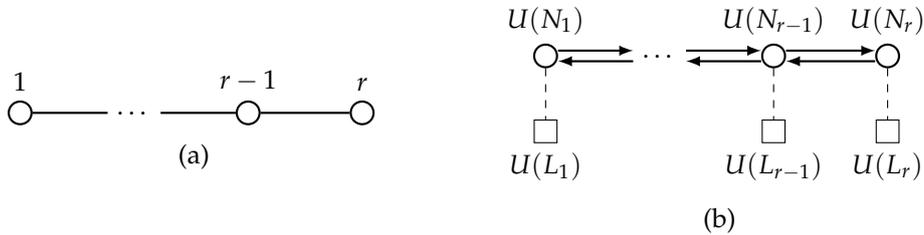
\begin{figure}
  \centering
  \begin{minipage}[c]{0.4\linewidth}
    \centering
    \begin{small}
      \begin{tikzpicture}
        \node (UN1) at (0,0) [circle, thick, draw, inner sep=2pt, minimum size=3mm, label=above:\( 1\) ] {};
        \node (UNr1) at (3,0) [circle, thick, draw, inner sep=2pt, minimum size=3mm, label=above:\( r-1 \) ] {};
        \node (UNr) at (4.5,0) [circle, thick, draw, inner sep=2pt, minimum size=3mm, label=above:\( r \) ] {};
        
        \node (dot) at (1.5,0) []{\dots};
        
        \draw (UN1) to (dot) [thick];
        \draw (dot) to (UNr1) [thick];
        \draw (UNr1) to (UNr) [thick];
      \end{tikzpicture}
    \end{small}
    \subcaption{}\label{fig:diagram-Ar-a}
  \end{minipage}
  \begin{minipage}[c]{0.5\linewidth}
    \centering
    \begin{small}
      \begin{tikzpicture}
        \node (UN1) at (0,0) [circle, thick, draw, inner sep=2pt, minimum size=3mm, label=above:\( U(N_1) \) ] {};
        \node (UNr1) at (3,0) [circle, thick, draw, inner sep=2pt, minimum size=3mm, label=above:\( U(N_{r-1}) \) ] {};
        \node (UNr) at (4.5,0) [circle, thick, draw, inner sep=2pt, minimum size=3mm, label=above:\( U(N_r) \) ] {};
        
        \node (UL1) at (0,-1) [draw, inner sep=2pt, minimum size=3mm, label=below:\( U(L_1) \) ] {};
        \node (ULr1) at (3,-1) [draw, inner sep=2pt, minimum size=3mm, label=below:\( U(L_{r-1}) \) ] {};
        \node (ULr) at (4.5,-1) [draw, inner sep=2pt, minimum size=3mm, label=below:\( U(L_r) \) ] {};
        
        \node (dot) at (1.5,0) []{\dots};
        
        \darrow{UN1}{dot}
        \darrow{dot}{UNr1}
        \darrow{UNr1}{UNr}
        
        \draw (UN1) to (UL1) [dashed];
        \draw (UNr1) to (ULr1) [dashed];
        \draw (UNr) to (ULr) [dashed];
      \end{tikzpicture}
    \end{small}
    \subcaption{}\label{fig:diagram-Ar-b}
  \end{minipage}
  \caption{Dynkin diagram (a) and quiver diagram (b) for the \( A_r \)
    series. In
    the quiver diagram, nodes are gauge groups and squares flavor groups.  Each arrow represents a bifundamental and each
    dotted line a pair fundamental--antifundamental. Each node carries also an adjoint
    field (not represented).}
  \label{fig:diagram-Ar}
\end{figure}

The Cartan matrix of $A_r$~series ($SU(r+1)$) has the form
\begin{equation}
  \label{sun}
  A = \begin{pmatrix}
    2 &-1 & 0 & \dots & & 0 \\
    -1 & 2 &-1 & \dots & & 0 \\
    0 & -1 & 2 &-1 & \dots & 0 \\
    \vdots & & \ddots & \ddots & \ddots & \vdots\\
    0 & 0 & \dots & -1 & 2 & -1 \\
    0 & 0 & \dots & 0 & -1 & 2 
  \end{pmatrix} \, ,
\end{equation}
which gives rise to the Dynkin diagram in Figure~\ref{fig:diagram-Ar-a}.
The $a$-th node of the Dynkin diagram translates directly to the $a$-th node of the quiver gauge theory.  Its color group  \( U(N_a) \) is determined by the superselection sector; the length parameter fixes the flavor group \( U(L_a) \) which is attached to the $a$-th node by fundamental and antifundamental fields $Q_k^a,\, \overline Q_k^a$, $k = 1,\dots L_a$.
Each arrow between nodes $a$ and $b$ of the quiver diagram corresponds to a bifundamental field $B^{a,b} $ of $U(N_a) \times U(N_{b})$. Moreover, each node can carry an adjoint field $\Phi^a$ (not pictured).

\paragraph{Brane construction.}

\begin{table}
  \centering
  \begin{tabular}{llcccccccccc}
    \toprule
    &  & 0 & 1 & 2 & 3 & 4 & 5 & 6 & 7 & 8 & 9 \\  \midrule 
    & fluxbrane & $\times $ & $\times $ & $\times $ & $\times $ &&&&&& $\times$ \\
    & NS5 & $\times $ & $\times $ & & & & & $\times $ & $\times $ & $\times $ & $\times$ \\
    & D2 & $\times $ & $\times $ & $\times $ & & & & & &  \\ \midrule
    \( \mathcal{N}=(2,2) \) & D4 & $\times $ & $\times $ & & $\times $ & $\times$ & $\times$  \\ \midrule
    \( \mathcal{N}=(1,1) \) & D4' & $\times $ & $\times $ & & $\times $ & & & & & $\times$ & $\times $ \\ \bottomrule
  \end{tabular}
  \caption{Type~\textsc{iia} Brane configuration for the 2d \( \mathcal{N}=(2,2) \) and \( \mathcal{N}=(1,1) \) theories living on the \D2 branes and reproducing the \( A \)~series symmetry. The crosses in the fluxbrane row mark the directions in which no identifications are imposed.}
  \label{tab:NS5-IIAembedding}
\end{table}

One can construct the $A_r$ quiver gauge theory in type~\textsc{iia} string theory by placing
$r+1$ parallel \NS5--branes extended in the $16789$~directions into a
flat background~\cite{Hanany:1996ie} (see Figure~\ref{fig:Dn-Branes}
and Table~\ref{tab:NS5-IIAembedding}). When the \NS5s coincide, the
six-dimensional \( \mathcal{N}=(2,0) \) type~\textsc{iia} theory acquires a \(
SU(r+1) \) symmetry that acts on the tensionless strings. Imposing
periodic boundary conditions for \( x_1 \), we can T--dualize and
obtain a system of coincident \NS5--branes in type~\textsc{iib}, which now has
standard \( SU(r+1) \) gauge symmetry%
\footnote{The periodicity of \( x_1 \) translates into periodic
  boundary conditions for the two-dimensional gauge theory. This is
  necessary in order to reproduce the Bethe Ansatz equations.}%
~\cite{Witten:1995zh}.  This is the enhanced \( SU(r+1) \) symmetry
that acts on the ground states of the effective theory of the probe
\D2--branes.

Suspended between the $a$-th and $(a+1)$-st \NS5--brane is a stack of
$N_a$ \D2--branes extended in the $12$~directions.  The motions of the open strings
within the $a$-th stack of \D2--branes in the
$67$~directions are described by the adjoint fields \( \Phi^a \), while
the strings going from the $a$-th to the $(a+1)$-st stack correspond to the
bifundamental fields \( B \).  
This configuration preserves eight
supersymmetries, which correspond to an \( \mathcal{N} = (4,4) \) gauge
theory in two dimensions. Gauge invariance and supersymmetry fix the
superpotential couplings of adjoints and bifundamentals to
\begin{equation}
  \label{eq:supo-BPhiB}
  W = \sum_{a=1}^{r-1} B^{a, a+1} \Phi^{a+1} B^{a+1,a} - B^{a+1,a} \Phi^a B^{a,a+1} \, .  
\end{equation}

The flavor groups live on stacks of $L_a$ D4--branes. They can be
added in two different ways:
\begin{enumerate}
\item \D4'--branes extended in the $1389$~directions break half of
  the supersymmetry and lead to an \( \mathcal{N} = (2,2) \) gauge
  theory on the \D2--branes. In this case, the D4--branes can be made to
  coincide with one of the \NS5 in the $x_2$~direction in
  which the \NS5--branes are separated, and broken in two. We can then
  distinguish \emph{upper-half} D4'--branes ($x_3>0$) and
  \emph{lower-half} D4'--branes ($x_3<0$). The strings from the
  D2--branes to the upper-half D4'--branes are described by fundamental
  fields \( Q^a_k \) and those going to the lower-half D4'--branes by
  antifundamental fields \( \overline Q^a_k \).
\item \D4--branes extended in the \( 1345 \)~directions preserve the
  same \( \mathcal{N} = (4,4) \) supersymmetry as the \NS5--\D2
  system. In this case, gauge invariance and supersymmetry require the
  presence of a superpotential term
  \begin{equation}
    \label{eq:N44-QPhiQ}
    W = \sum_{k=1}^{L_a} Q_k^a \Phi^a \overline Q_k^a.    
  \end{equation}
\end{enumerate}

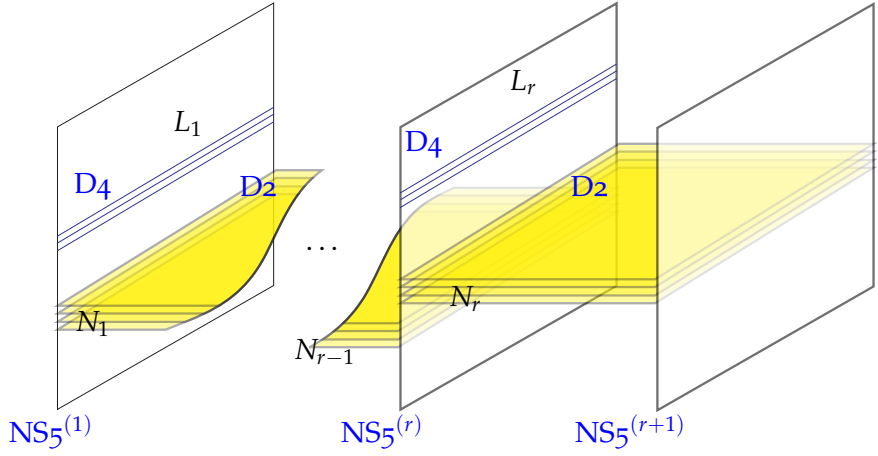
\begin{figure}
  \centering
  \begin{tikzpicture}%
    \node[anchor=south west,inner sep=0] (image) at (0,0){\includegraphics[height=5.5cm]{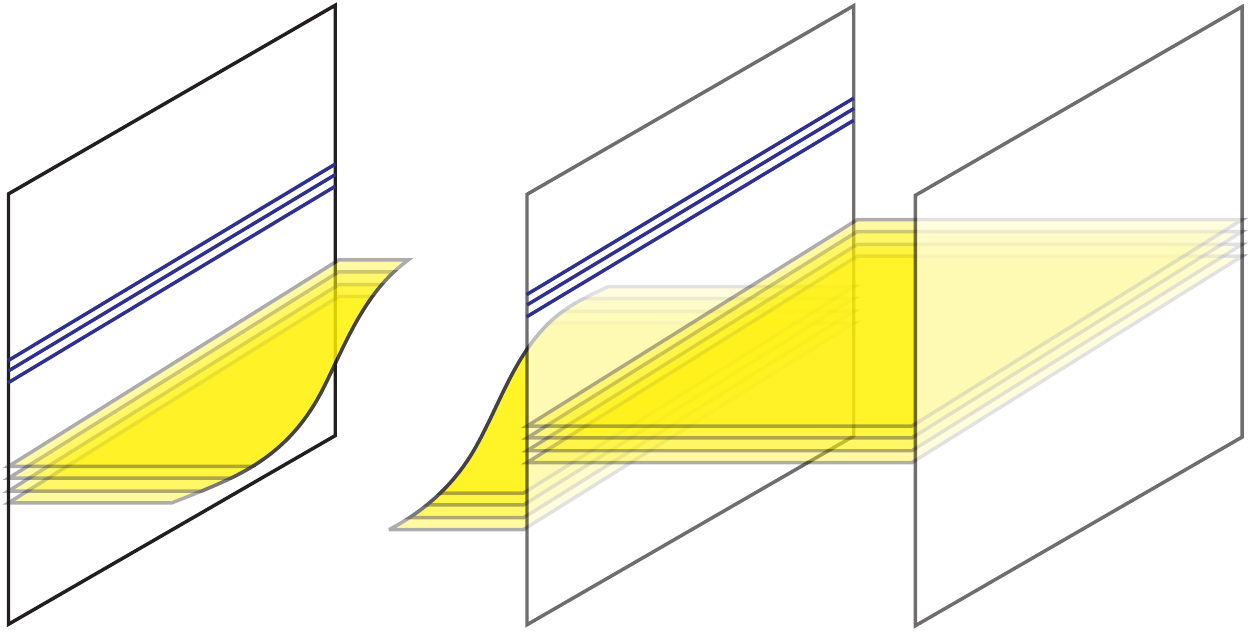}};
  \begin{scope}[x={(image.south east)},y={(image.north west)}]

    \node at (0.33,0.4) {\dots};
    \node[blue] at (0,-.04) {$\NS5^{(1)}$};
    \node[blue] at (0.4,-.04) {$\NS5^{(r)}$};
    \node[blue] at (0.7,-.04) {$\NS5^{(r+1)}$};

    \node[blue] at (.05, .55){\D4};
    \node[blue] at (.25, .55){\D2};
    \node[blue] at (.45, .65){\D4};
    \node[blue] at (.65, .55){\D2};

    \node at (0.05,0.22) {\( N_1 \)}; 
    \node at (0.33,0.15) {\( N_{r-1} \)}; 
    \node at (0.5,0.28) {\( N_{r} \)}; 

    \node at (0.165,0.7) {\( L_{1} \)}; 
    \node at (0.57,0.8) {\( L_{r} \)};

  \end{scope}

  \end{tikzpicture}
  \caption{Brane cartoon for the \( A_r \) model}
  \label{fig:Dn-Branes}
\end{figure}

\paragraph{Twisted masses.}

To reproduce the parameters of the integrable system, \emph{twisted
  masses} need to be turned on for all fields, which reduces the
supersymmetry of the system by one half.

The twisted masses of the adjoint and bifundamental fields do
\emph{not} originate from the brane construction, but can be turned on
by placing the above construction into a so-called \emph{fluxtrap
  background}~\cite{Hellerman:2011mv}, which we will detail in the
following. For the moment we observe that the superpotential term
coupling bifundamentals and adjoints in Equation~\eqref{eq:supo-BPhiB} preserves a \( U(1) \) symmetry
\begin{equation}
  \eu^{\imath m} : ( B^{a+1,a}, \Phi^a, B^{a,a+1} ) \mapsto ( \eu^{-\imath m/2 }B^{a+1,a}, \eu^{\imath m} \Phi^a, \eu^{-\imath m/2} B^{a,a+1} ) \, ,
\end{equation}
which constrains the corresponding twisted masses to obey
\begin{equation}
  \label{eq:bifundamental-mass-equation}
  \wt{m}^{\bif}{}^{(a+1,a)} + \wt{m}^{\adj}{}^{(a)} + \wt{m}^{\bif}{}^{(a,a+1)} = 0 \, .
\end{equation}
A twisted mass \( m \) for the adjoints \( \Phi^a \) 
automatically requires twisted masses for the bifundamentals that can
be gauge-fixed to \( -m/2 \).

\bigskip

For the fundamentals and antifundamentals, the two cases above must again be
distinguished.
\begin{enumerate}
\item In the case of \( \mathcal{N} = (2,2) \) supersymmetry broken to
  \( \mathcal{N} = (1,1) \), twisted masses can be turned on by moving
  the upper-half and lower-half \D4'--branes away from each
  other. Their $x_6 + \imath x_7$--values are the twisted masses for
  the $Q_k^a,\, \overline Q_k^a$--fields~\cite{Hanany:1997vm}. In
  particular, the \emph{relative separation} of the upper-half and
  lower-half \D4' corresponds to the $\Lambda^a_k$~parameter from the
  spin chain, while the \emph{center of mass} of the position
  corresponds to the $\nu^a_k$~parameter.
\item In the case of \( \mathcal{N} = (4,4) \) supersymmetry broken to
  \( \mathcal{N} = (2,2) \), the representations of the symmetry group
  at each position of the chain is restricted to $\Lambda^a_k = 1$ or
  \( \Lambda^a_k = 0 \), while it is impossible to turn on impurities,
  thus $\nu^a_k=0$. The twisted masses of the fundamental and
  antifundamental fields are inherited directly from the fluxtrap
  background via the superpotential term $\sum_{k=1}^{L_a} Q_k^a
  \Phi^a \overline Q_k^a$ that we have encountered in
  Equation~\eqref{eq:N44-QPhiQ}, following the same mechanism outlined
  above for the twisted masses of the bifundamentals
  (see~\cite{Hellerman:2011mv}).
\end{enumerate}

\paragraph{The Fluxbrane Background}

Let us concentrate on the \NS5--\D2 system. In order to understand the
fluxtrap background and the corresponding twisted masses for the
fields \( \Phi \) and \( B \) it is convenient to consider the T--dual
type~\textsc{iib} setup in which \NS5s are extended in the \( 16789 \)
directions and \D3--branes are extended in the \( 128 \) directions.

The fluxbrane construction is obtained by imposing a set of
identifications \( \Gamma \)  on the space orthogonal to the \D3--branes. The
directions \( 3 \) and \( 9 \) will be spectators. Let \( x_8 \) be
periodic with radius \( \wt R \),
\begin{equation}
  x_8 = \wt R \wt u \,, \hspace{2em} \wt u \simeq \wt u + 2 \pi k_1 \, ,  
\end{equation}
and the remaining four directions \( 4567 \) be described by
cylindrical coordinates \( (\rho_1, \theta_1, \rho_2, \theta_2 ) \).
The space $\setR^5 / \Gamma $ is obtained by imposing 
\begin{equation}
  \label{eq:theta-periodicity}
  \begin{cases}
    \wt u \simeq \wt u + 2 \pi \, k_1 \, , \\
    \theta_1 \simeq \theta_1 + 2 \pi m \wt R \, k_1 \, , \\
    \theta_2 \simeq \theta_2 - 2 \pi m \wt R \, k_1 \, ,
  \end{cases} \hspace{2em} k_1 \in \setZ \, ,
\end{equation}
in addition to the standard identifications for cylindrical coordinates,
\begin{align}
  \theta_1 &\simeq \theta_1 + 2 \pi \, k_2 \, , & \theta_2 &\simeq
  \theta_2 + 2 \pi \, k_3 \, , \hspace{2em} k_2, k_3 \in \setZ \, .
\end{align}
In order to disentangle the periodicities we introduce the new angular variables
\begin{equation}
  \begin{cases}
    \phi_1 = \theta_1 - m \wt R \wt u \, ,\\
    \phi_2 = \theta_2 + m \wt R \wt u  \, ,
  \end{cases}
\end{equation}
which are \( 2 \pi \)-periodic.

Since the \NS5s are much heavier than the \D3--branes, we can treat the
latter as probes propagating in the background generated by the
backreaction of the former.  In rectilinear coordinates ($x_4 + \imath
x_5 = \rho_1 \eu^{\imath \phi_1}, x_6 + \imath x_7 = \rho_2
\eu^{\imath \phi_2}$) the fields in the bulk read:
\begin{align}
  \begin{split}
    \wt {\di s}^2 &= - \di x_0^2 + \di x_1^2 + \di x_8^2 + \di x_9^2 + U \left[ \di x_2^2 + \di x_3^2 + \sum_{i=4}^5
      \left( \di x_i + m V^i \di x_8 \right)^2 \right] \\
    & \phantom{{}={}} + \sum_{i=6}^7
    \left( \di x_i + m V^i \di x_8 \right)^2   \, , 
  \end{split} \\
  \di B &= * \di U = b_i \di x^i \wedge \left( \di \phi_1 + m \wt R
    \di \wt u \right),  \\
  \Phi &= \log(\wt R\,g_3^2)+ \frac{1}{2}\log U \,,
\end{align}
where
\begin{align}
  U &= 1 + \sum_{a=1}^{r+1} \frac{ \alpha'}{\left( x_2 - X_2^{(a)}
    \right)^2 + \left( x_3 - X_3^{(a)} \right)^2 + x_4^2 + x_5^2} \, , \\
  V^i \del_i &= - x^5 \del_{x_4} + x^4 \del_{x_5} + x^7 \del_{x_6} -
  x^6 \del_{x_7} = \del_{\phi_1} - \del_{\phi_2} \, ,
\end{align}
and \( (X_2^{(a)}, X_3^{(a)}) \) is the position of the \( a \)-th
\NS5.  This provides the \emph{$\Omega$--deformation of the NS5
  background}. Note that the \NS5--branes are forced to sit in \(
X^{(a)}_3 = X^{(a)}_4 = 0 \) in order to preserve the rotation
symmetry in the \( (x_4, x_5) \) plane that is necessary for the
fluxbrane construction.

\bigskip

Now we can T--dualize in the direction $\wt u$ to go back to our original \NS5--\D2 configuration. 
The bulk fields in the \NS5--fluxtrap background are given by 
\begin{align}
  \begin{split}
    \di s^2 &= - \di x^2_0 + \di x_1^2 + U \, \left[ \di x_2^2 + \di
      x_3^2 + \di x_4^2 + \di x_5^2 \right] + \di x_6^2 + \di x_7^2 +
    \di x_9^2\,,  \\
    & \phantom{{}={}} + \frac{1}{\Delta^2} \left[ \left( m \, b_i \di
        x^i + \di x_8 \right)^2 - m^2 \left( U \left( x_4 \di x_5 -
          x_5 \di x_4 \right) - x_6 \di x_7 + x_7 \di x_6 \right)^2
    \right],
\end{split} \\
    B &= \frac{1}{\Delta^2} \left[ b_i \di x^i \wedge
      \left( \di \phi_1 + m^2 \rho_2^2 \left( \di \phi_1 + \di \phi_2 \right) \right)
   + m %
   \left( U \, \rho_1^2 \di \phi_1 - \rho_2^2 \di
     \phi_2 \right)%
   \wedge \di x_8 \right]\,, \\
  \eu^{-\Phi} &= \frac{1}{g_3^2 \sqrt{\alpha'}} \frac{\Delta}{\sqrt{U}} %
\, ,
\end{align}
where
\begin{equation}
  \Delta^2 = 1 + m^2 \left( U  \rho_1^2 + \rho_2^2
  \right) %
  \, .
\end{equation}
The \NS5--fluxtrap background preserves eight supercharges. The conserved Killing spinors have been found in~\cite{Hellerman:2011mv} and read
\begin{equation}
  \begin{cases} 
      \epsilon_L = \eu^{-\Phi/8}
      \left( \Id + \Gamma_{11} \right) \proj{NS5}_-
      \proj{flux}_- \exp [\tfrac{1}{2} \phi_1 \Gamma_{45} +
      \tfrac{1}{2} \phi_2 \Gamma_{67} + \tfrac{1}{2} \phi_3 \Gamma_{23} ] \epsilon_0\,, \\
      \epsilon_R = \eu^{-\Phi/8} \left( \Id - \Gamma_{11} \right)
      \Gamma_u \proj{NS5}_+ \proj{flux}_- \exp[ \tfrac{1}{2} \phi_1
      \Gamma_{45} + \tfrac{1}{2} \phi_2 \Gamma_{67} + \tfrac{1}{2}
      \phi_3 \Gamma_{23}] \epsilon_1\,,
  \end{cases}
\end{equation}
where \( \epsilon_0 \) and \( \epsilon_1 \) are constant Majorana
spinors, 
\begin{equation}
  \Gamma_u =  \frac{m \rho_1}{\Delta} \Gamma_5 - \frac{m \rho_2}{\Delta} \Gamma_7 +
  \frac{1}{\Delta} \Gamma_8 \, ,
\end{equation}
and
\begin{align}
  \label{eq:susy-projectors}
  \proj{flux}_\pm &= \tfrac{1}{2} \left( \Id \pm \Gamma_{4567} \right) \, , & \proj{\NS5}_\pm &= \tfrac{1}{2} \left( \Id \pm \Gamma_{2345} \right) \, .
\end{align}

\paragraph{Other parameters.}

At this point we can identify the geometric origin of the remaining
gauge theory parameters.
\begin{itemize}
\item The gauge couplings \( e_a \) for the \( U(N_a) \) theories are given by
  the separations of the \NS5--branes in the \( x_2 \) direction (the
  direction in which the \D2s have a finite extension):
  \begin{equation}
    \label{eq:gauge-coupling}
    \left( X_2^{(a+1)} - X_2^{(a)} \right) \frac{\ell_{\text{st}}}{g_{\text{st}}} = \frac{1}{e_a^2} \, ,
  \end{equation}
  where \( \ell_{\text{st}} \) is the string length and \(
  g_{\text{st}} \) is the type~\textsc{iia} string coupling constant.
\item The Fayet--Iliopoulos (\textsc{fi}) parameters are proportional
  to the separations of the \NS5--branes in the \( x_3 \)
  direction\footnote{In the \( \mathcal{N} = (4,4) \) case broken to
    \( \mathcal{N}=(2,2) \), the directions \( 3, 4 \text{ and } 5\)
    are initially equivalent, which corresponds to the three
    independent \textsc{fi} parameters of the \( \mathcal{N} = (4,4)
    \) theory. This symmetry is broken by the fluxtrap that fixes the
    \( x_4 \text{ and } x_5 \) positions of the \NS5s to the origin, thus leaving
    the single \textsc{fi} parameter of \( \mathcal{N} = (2,2) \).}:
  \begin{equation}
    \frac{X_3^{(a+1)} - X_3^{(a)}}{\ell_{\text{st}} g_{\text{st}}} = - \tau_a \, .       
  \end{equation}
\end{itemize}
The \( \theta \) angle does not have a direct geometric interpretation
in type~\textsc{iia}, but requires the uplift to M-theory.
 
\bigskip

Both a non-vanishing gauge coupling and a non-vanishing \textsc{fi}
parameter break the \( A_r \) symmetry, albeit in very different ways.
\begin{itemize}
\item When the gauge coupling vanishes, we conjecture that all the
  massive states decouple, thus leaving only the \( A_r \)-charged
  ground states\footnote{This argument does not exclude the presence
    of \( A_r \)-charged \textsc{bps}-protected states, whose
    existence is currently under investigation.}. The new massive
  states appearing for non-zero gauge coupling do not have an analogue
  on the integrable system side. This is why the coupling \( e_a \) in
  Equation~\eqref{eq:gauge-coupling} does not correspond to a
  parameter of the spin chain.
\item A non-vanishing \textsc{fi} parameter does not add new states,
  but breaks the \( A_r \) symmetry. This corresponds directly to
  imposing a non-periodic boundary condition on the chain. 
  The Hilbert space remains the same, but the Hamiltonian
  does not commute anymore with the generators of the global \( A_r
  \). The \textsc{fi} parameter acts as a \emph{twist} that splits up
  the energy levels of all \( A_r \) multiplets
  (see~\cite{Kulish:1997dy}).
\end{itemize}

\paragraph{Ground states as BPS states.}

The ground states of the \D2 probe branes extended in the directions \( 1 \text{ and } 2 \) can be understood as \( 1/2 \)--\textsc{bps} objects in the \NS5--fluxbrane background (four supercharges).
As expected, the effect of the fluxtrap is that the minimum energy configurations are localized at the bottom of a potential at \( x_3 = x_4 = x_5 = 0 \). 
More general \textsc{bps} states with less supersymmetry can be constructed, but in general they are not localized and hence non-normalizable. 
A \textsc{dbi} analysis around the stable configurations shows that in the effective gauge theory description, all the adjoint fields \( \Phi^a \) which describe the motion in the directions \( 6 \text{ and } 7 \) receive the same twisted mass term \( m \).
This mass is propagated to the bifundamental fields \( B^{a,a+1} \) via the superpotential coupling as explained above.

\bigskip

We have thus constructed a brane realization of the full $A_r$ quiver gauge theory which includes all the parameters of the corresponding integrable model, in particular the twisted masses. The symmetry group of the spin chain is reflected in the symmetry of the background containing the parallel \NS5--branes. The Gauge/Bethe correspondence implies the action of the background $A_r$ symmetry on the probe branes such that the raising and lowering operators \( E^\pm_a  \) change the gauge group at the \( a \)-th node:
\begin{equation}
  \label{eq:raising-lowering-N}
  E^\pm_a : U( N_a ) \mapsto U (N_a \pm 1 ) \, .
\end{equation}

\section{The $D$~series}
\label{sec:d-series}

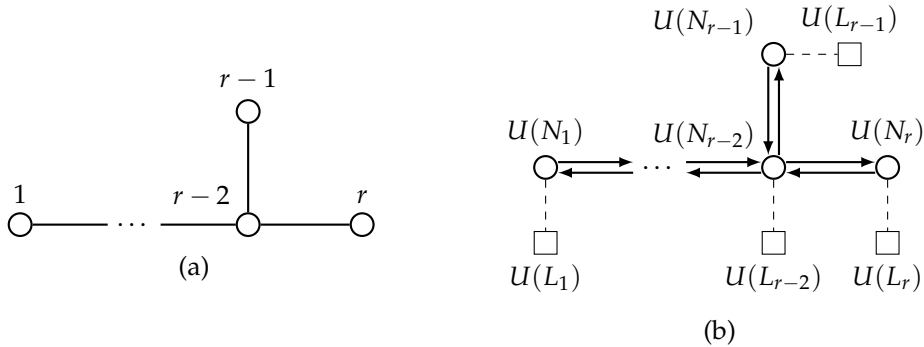
\begin{figure}[b]
  \centering
  \begin{minipage}[c]{0.4\linewidth}
    \centering
    \begin{small}
      \begin{tikzpicture}
        \node (UN1) at (0,0) [circle, thick, draw, inner sep=2pt, minimum size=3mm, label=above:\( 1 \) ] {};
        \node (UNr2) at (3,0) [circle, thick, draw, inner sep=2pt, minimum size=3mm, label=above left:\( r-2  \) ] {};
        \node (UNr1) at (3,1.5) [circle, thick, draw, inner sep=2pt, minimum size=3mm, label=above:\( r-1 \) ] {};
        \node (UNr) at (4.5,0) [circle, thick, draw, inner sep=2pt, minimum size=3mm, label=above:\( r \) ] {};

        \node (dot) at (1.5,0) []{\dots};
        
        \draw (UN1) to (dot) [thick];
        \draw (dot) to (UNr2) [thick];
        \draw (UNr2) to (UNr1) [thick];
        \draw (UNr2) to (UNr) [thick];
        
      \end{tikzpicture}
    \end{small}    
    \subcaption{}\label{fig:diagram-Dr-a}
  \end{minipage}
  \begin{minipage}[c]{0.5\linewidth}
    \centering
    \begin{small}
      \begin{tikzpicture}
        \node (UN1) at (0,0) [circle, thick, draw, inner sep=2pt, minimum size=3mm, label=above:\( U(N_1) \) ] {};
        \node (UNr2) at (3,0) [circle, thick, draw, inner sep=2pt, minimum size=3mm, label=above left:\( U(N_{r-2}) \) ] {};
        \node (UNr1) at (3,1.5) [circle, thick, draw, inner sep=2pt, minimum size=3mm, label=above left:\( U(N_{r-1}) \) ] {};
        \node (UNr) at (4.5,0) [circle, thick, draw, inner sep=2pt, minimum size=3mm, label=above:\( U(N_r) \) ] {};
        
        \node (UL1) at (0,-1) [draw, inner sep=2pt, minimum size=3mm, label=below:\( U(L_1) \) ] {};
        \node (ULr2) at (3,-1) [draw, inner sep=2pt, minimum size=3mm, label=below:\( U(L_{r-2}) \) ] {};
        \node (ULr1) at (4,1.5) [draw, inner sep=2pt, minimum size=3mm, label=above:\( U(L_{r-1}) \) ] {};
        \node (ULr) at (4.5,-1) [draw, inner sep=2pt, minimum size=3mm, label=below :\( U(L_r) \) ] {};
        
        \node (dot) at (1.5,0) []{\dots};
        
        \darrow{UN1}{dot}

        \darrow{dot}{UNr2}

        \darrowv{UNr2}{UNr1}
        \darrow{UNr2}{UNr}
        
        \draw (UN1) to (UL1) [dashed];
        \draw (UNr2) to (ULr2) [dashed];
        \draw (UNr1) to (ULr1) [dashed];
        \draw (UNr) to (ULr) [dashed];
        
      \end{tikzpicture}
    \end{small}
    \subcaption{}\label{fig:diagram-Dr-b}
  \end{minipage}
  \caption{Dynkin diagram and quiver diagram for \( D_r \) series. In
    the quiver diagram, nodes are gauge groups and squares flavor groups.  Each arrow represents a bifundamental and each
    dotted line a pair fundamental--antifundamental. Each node carries
    also an adjoint field (not represented).}
  \label{fig:diagram-Dr}
\end{figure}

The Cartan matrix of the $D_r$~series ($SO(2r)$) has the form
\begin{equation}
  \label{sonn}
  A = \begin{pmatrix}
    2 & -1 & 0 & \dots & & & 0 \\
    -1 & 2 & -1 & \dots & & & 0 \\
    0 & -1 & 2 & -1 & & & 0 \\
    \vdots & & \ddots & \ddots & \ddots & & \vdots \\
    0 & 0 & \dots & & 2 & -1 & -1 \\
    0 & 0 & \dots & & -1 & 2 & 0 \\
    0 & 0 & \dots & & -1 & 0 & 2 
  \end{pmatrix} \, ,
\end{equation}
which gives rise to the Dynkin diagram in Figure~\ref{fig:diagram-Dr-a}.

Looking at the Dynkin diagram, we see that up to the second last node,
it is the same as the one of the $A_r$~series. Up to this node we
can therefore replicate the brane construction of the last section,
consisting of parallel NS5--branes with stacks of D2--branes between
them. Let us concentrate on the nodes \( r - 2, r - 1 \text{ and } r
\). From the Dynkin diagram, we learn that there should be no
bifundamental fields between the last two nodes, but extra
bifundamental fields between nodes $r-2$ and $r$. To achieve this, we
need to make use of a somewhat more exotic object, which is best
described as the S--dual of a D5--brane coincident with an
O5--plane~\cite{Sen:1996na, Sen:1998ii, Kapustin:1998fa,
  Hanany:1999sj, Hellerman:2005ja}. For further convenience, we will
name this object NO5. In our picture, it corresponds to a
$\mathbb{Z}_2$ orbifold $\mathcal{I}_4$ which reflects the $2345$~directions times the action of $(-1)^{F_L}$, where $F_L$
is the left-moving fermion number. The \NO5 orbifold preserves exactly
the same supersymmetries as an \NS5--brane extended in the \( 16789 \)
directions as discussed in Appendix~\ref{sec:supersymm-orbif}.

\paragraph{Field content.}

As for the fluxbrane construction, it is convenient to start from the T--dual configuration in
which \D3--branes are suspended between \NS5--branes and the \NO5 orbifold%
\footnote{The \NO5 remains invariant under T--duality in a parallel direction~\cite{Kapustin:1998fa}.}.
Consider a stack of \( N_2 + N_3 \) \D3--branes extended in the \( 128 \) directions, suspended between the \NO5--plane localized at \( x_2 = x_3 = x_4 = x_5 = 0 \) and an \NS5--brane localized at \( (x_2 = X_2^{(2)}, x_3 = X_3^{(2)}, x_4 = 0, x_5 = 0) \) (see Table~\ref{tab:NO5-IIBembedding}).

\begin{table}
  \centering
  \begin{tabular}{llcccccccccc}
    \toprule
    &  & 0 & 1 & 2 & 3 & 4 & 5 & 6 & 7 & 8 & 9 \\  \midrule 
    & NS5 & $\times $ & $\times $ & & & & & $\times $ & $\times $ & $\times $ & $\times$ \\
    & NO5 & $\times $ & $\times $ & & & & & $\times $ & $\times $ & $\times $ & $\times$ \\
    & D3 & $\times $ & $\times $ & $\times $ & & & & & & $\times$ \\ \midrule

    \( \mathcal{N}=4 \) & D5 & $\times $ & $\times $ & & $\times $ & $\times$ & $\times$ &
    & & $\times $ & \\ \midrule
   \( \mathcal{N}=2 \) & D3' & $\times $ & $\times $ & & $\times $ & & & & &  & $\times $ 
   \\ \bottomrule
  \end{tabular}
  \caption{Type~\textsc{iib} brane configuration for the 3d \( \mathcal{N}=4 \)
    and \( \mathcal{N}=2 \) theories living on the \D3--branes. The crosses
    in the \NO5 orbifold row mark the directions in which the space
    remains flat and no identifications are imposed.}
  \label{tab:NO5-IIBembedding}
\end{table}

The six-dimensional theory living on the \NO5 orbifold has \( SO(2) \)
symmetry, under which the \D3--branes are magnetically charged. This
means that in the stack of \D3s we can distinguish the \( N_3 \)
positively charged $\D3^+$ from the \( N_2 \)  negatively
charged $\D3^-$. The \( U(N_2 + N_3) \) symmetry
is broken to \( U(N_2) \times U(N_3) \), as one can verify by writing
the partition function for open strings between two \D3--branes:
strings between two branes with the same \( SO(2) \) charge and
strings between \D3${}^+$ and \D3${}^-$ behave differently under the
inversion of coordinates and do not mix. From the effective gauge
theory point of view, the oscillations of the \D3--branes transverse
to the \NO5 result in adjoint fields for both \( U(N_2) \) and \(
U(N_3) \), while the bifundamentals are projected out by the action of
\( \mathcal{I}_4 \) (see~\cite{Kapustin:1998fa}).  

At this point we can add a second parallel \NS5--brane at \( x_2 =
X_2^{(1)}, x_3 = X_3^{(1)} \) and a stack of \( N_1 \) \D3--branes
suspended between the two \NS5--branes. The configuration around the
\NS5 at \( (X_2^{(2)}, X_3^{(2)}) \) is the same that we have
encountered in the previous section. This means that together with the
adjoints of \( U(N_1) \) coming from the \D3--branes on the left, we
also have bifundamentals for \( U(N_1) \times U(N_2) \) and \( U(N_1)
\times U(N_3) \) which are coupled via a \( \sum_{a=2}^3 (B^{a,1} \Phi^1 B^{1,a} -
B^{1,a} \Phi^a B^{a,1}) \) term. The field content of the theory is
thus summarized by a quiver diagram with \( D_3 \) shape (see
Figure~\ref{fig:diagram-Dr-b}). Higher \( D_r \) diagrams are obtained
by adding more parallel \NS5s with \D3--branes suspended in between
(see the cartoon in Figure~\ref{fig:Dn-Branes}).

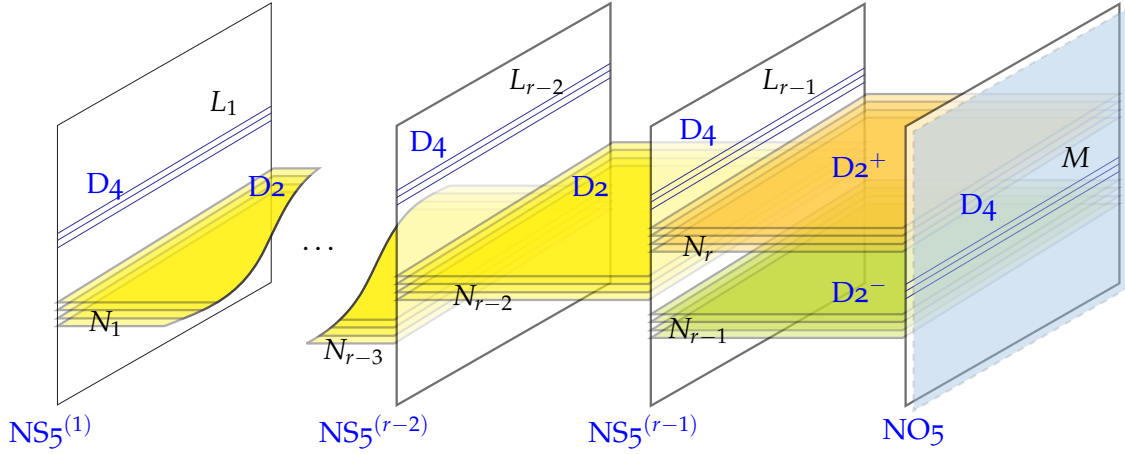
\begin{figure}
  \centering
  \begin{tikzpicture}%
    \node[anchor=south west,inner sep=0] (image) at (0,0){\includegraphics[height=5.5cm]{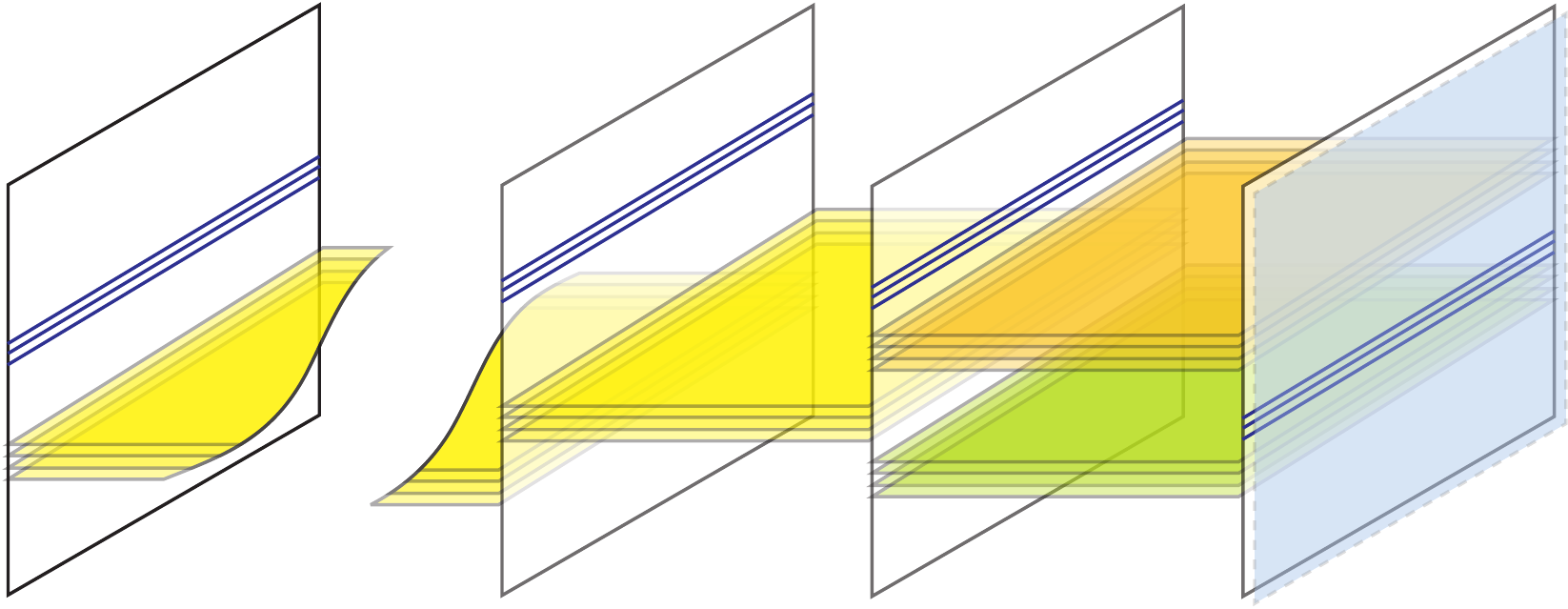}};
  \begin{scope}[x={(image.south east)},y={(image.north west)}]

    \node at (0.25,0.4) {\dots};
    \node[blue] at (0,-.04) {$\NS5^{(1)}$};
    \node[blue] at (0.3,-.04) {$\NS5^{(r-2)}$};
    \node[blue] at (0.55,-.04) {$\NS5^{(r-1)}$};
    \node[blue] at (0.8,-.04) {\NO5};

    \node[blue] at (.05, .55){\D4};
    \node[blue] at (.2, .55){\D2};
    \node[blue] at (.35, .65){\D4};
    \node[blue] at (.5, .55){\D2};
    \node[blue] at (.6, .68){\D4};
    \node[blue] at (.86, .5){\D4};

    \node[blue] at (.75, .6){\D2${}^+$};
    \node[blue] at (.75, .3){\D2${}^-$};

    \node at (0.05,0.22) {\( N_1 \)}; 
    \node at (0.28,0.15) {\( N_{r-3} \)}; 
    \node at (0.4,0.28) {\( N_{r-2} \)}; 
    \node at (0.6,0.2) {\( N_{r-1} \)}; 
    \node at (0.6,0.4) {\( N_{r
} \)}; 

    \node at (0.16,0.75) {\( L_{1} \)}; 
    \node at (0.45,0.8) {\( L_{r-2} \)}; 
    \node at (0.685,0.8) {\( L_{r-1} \)}; 
    \node at (0.95,0.62) {\( M \)}; 

  \end{scope}

  \end{tikzpicture}
  \caption{Brane cartoon for the \( D_r \) model}
  \label{fig:Dn-Branes}
\end{figure}

Following the construction presented in the previous section, we can
add fundamentals to the two-dimensional gauge theory by inserting more branes into the system in two
ways:
\begin{enumerate}
\item breaking half of the supersymmetry such that the effective theory
  has \( \mathcal{N}=2 \) supersymmetry in three dimensions (four
  supercharges). In this case we add a stack of \D3'--branes
  extended in the \( 139 \) directions that end on the \NS5 and can be
  split into an upper and lower part (extended in \( x_3 \gtrless 0
  \));
\item respecting the same supersymmetries preserved by the
  \NO5--\NS5--\D3 system. In this case, the effective gauge theory has
  \( \mathcal{N}=4 \) supersymmetry in three dimensions (eight
  supercharges). This corresponds to having stacks of \D5--branes
  extended in the \( 13458 \)~directions and living in the intervals
  between the \NS5--branes.
\end{enumerate}
The new possibility with respect to the \( A_r \) configuration is
that a stack of \( M \) \D3'--branes or \D5--branes can end on the
\NO5 plane. In this case, the corresponding open strings give rise to \(
L_r = 2M + L_{r-1} \) fundamentals for the \( U(N_r) \) node
(see~\cite{Hanany:1999sj}).

\paragraph{Fluxbrane, fluxtrap and twisted masses.}

Now that we know that the gauge theory describing the oscillations of
D3--branes in the \NO5--\NS5--\D5 system has the right matter content
to reproduce the \( D \)~series Bethe Ansatz equations, we need to verify
that the orbifold action is compatible with the fluxbrane construction
that gives rise to the necessary twisted masses.

Following the previous section we start from the T--dual type~\textsc{iib} \NO5--NS5
system and introduce polar coordinates \( (\rho_1, \theta_1, \rho_2,
\theta_2) \) for four of the directions perpendicular to the
\D3--branes.  The two identifications that we want to impose are
\begin{itemize}
\item the \emph{orbifold action} which reflects \( \rho_1 \eu^{\imath
    \theta_1} \), leaving \( \rho_2 \eu^{\imath \theta_2} \)
  invariant, and reflects \( x_2, x_3 \), which are spectators for the
  fluxbrane:
  \begin{equation}
    \mathcal{I}_4 : ( x_2, x_3, \rho_1, \theta_1, \rho_2, \theta_2 ) \mapsto ( - x_2, - x_3, \rho_1, \theta_1 + \pi, \rho_2, \theta_2 ) \, .
  \end{equation}
\item the \emph{fluxbrane identifications} on flat space,
  \begin{equation}
    \begin{cases}
      \wt u \simeq \wt u + 2 \pi k_1\,, \\ \theta_1 \simeq \theta_1 + 2\pi
      m \wt R k_1\,, \\ \theta_2 \simeq \theta_2 - 2\pi m \wt R k_1 \, ,
    \end{cases} 
  \end{equation}
  which are disentangled by introducing the angles
  \begin{align}
    \phi_1 &= \theta_1 - m \wt R \wt u\,, & \phi_2 &= \theta_2 + m \wt R
    \wt u \, .
  \end{align}
\end{itemize}
We need to verify that it is possible to define
the action of \( \mathcal{I}_4 \) on the coordinates \( \phi_i \),
which automatically implement the fluxbrane identifications.  Using
the definition of \( \phi_1 \) and \( \phi_2 \), it is immediate to
see that the orbifold action remains simple,
\begin{equation}
  \mathcal{I}_4 :  ( x_2, x_3, \rho_1, \phi_1, \rho_2, \phi_2, \wt u ) \mapsto ( - x_2, - x_3, \rho_1, \phi_1 + \pi, \rho_2, \phi_2, \wt u ) \, .
\end{equation}
We conclude thus that the orbifold action is compatible with the fluxbrane
construction and it is formally the same as in flat space when written
for the disentangled coordinates.

At this point we can T--dualize the system in the direction \( x_8 \) and obtain the \NO5--\NS5--fluxtrap background in which the movement of the \D2--branes is described by a \( \mathcal{N}=(2,2) \) two--dimensional gauge theory with a twisted mass term \( m \) for the adjoints.  
Under T--duality, both the \D3'--branes and the \D5--branes turn into \D4--branes.  
In the \( \mathcal{N}=(2,2) \) case, the respective fundamentals and antifundamentals acquire a twisted mass \( -m/2 \) via the superpotential coupling as in Equation~\eqref{eq:bifundamental-mass-equation}. 
In the \( \mathcal{N}=(1,1) \) case, generic twisted masses can be obtained by breaking the \D4'--branes into an upper and a lower part (see the configuration outlined in Table~\ref{tab:NO5-IIAembedding}).  
The identification of the remaining parameters is the same as in the previous section.

Once more, the Gauge/Bethe correspondence implies an action of the \( SO(2r) \) group on the quiver gauge theories that we have realized. Raising and lowering operators \( E^\pm_a \) change the number of colors at the corresponding nodes  by one as in Equation~\eqref{eq:raising-lowering-N}.

\begin{table}
  \centering
  \begin{tabular}{llcccccccccc}
    \toprule
    &  & 0 & 1 & 2 & 3 & 4 & 5 & 6 & 7 & 8 & 9 \\  \midrule 
    & fluxbrane & $\times $ & $\times $ & $\times $ & $\times $ &&&&&& $\times$ \\
    & \NS5 & $\times $ & $\times $ & & & & & $\times $ & $\times $ & $\times $ & $\times$ \\
    & \NO5 & $\times $ & $\times $ & & & & & $\times $ & $\times $ & $\times $ & $\times$ \\
    & \D2 & $\times $ & $\times $ & $\times $ & & & & & &  \\ \midrule
    \( \mathcal{N}=(2,2) \) & \D4 & $\times $ & $\times $ & & $\times $ & $\times$ & $\times$  \\ \midrule
    \( \mathcal{N}=(1,1) \) & \D4' & $\times $ & $\times $ & & $\times $ & & & & & $\times$ & $\times $ \\ \bottomrule
  \end{tabular}
  \caption{Type~\textsc{iia} Brane configuration for the 2d \( \mathcal{N}=(2,2) \) and \( \mathcal{N}=(1,1) \) theories living on the \D2--branes and reproducing the \( D \)~series symmetry. The crosses in the fluxbrane and \NO5 rows mark the directions in which no identification is imposed.}
  \label{tab:NO5-IIAembedding}
\end{table}

\section{Dictionary and Outlook}

In this note we have extended the brane construction of the gauge system corresponding to the \textsc{xxx}\( {}_{1/2} \) model~\cite{Hellerman:2011mv} to the general case of \( A_r \text{ and } D_r \) spin chains.
The main novelty of this construction is the inclusion of twisted masses into the quiver gauge theories.  
The twisted masses introduce a potential for the ground states which can be identified -- via the Gauge/Bethe correspondence -- with the eigenvectors of a spin chain.
Our main result is that the enhanced background symmetry of the \NS5 (or \NO5--\NS5) system acts on the states of the gauge theories that describe the motion of the probe \D2--branes.

\bigskip

The results of the previous two sections are summarized in a dictionary, see Table~\ref{tab:dictionary}. 
It is based on the Gauge/Bethe dictionary table given in~\cite{Orlando:2010uu}, with a new column containing the translation to the parameters of the brane constructions for the $A_r$ and $D_r$ series.
As already discussed, we are indeed able to \emph{match all the parameters} of the integrable system/gauge theory to parameters in the string theory construction.  
Note that in Table~\ref{tab:dictionary} we assume the more general case of $\mathcal{N}=(1,1)$ supersymmetry, where the D4--branes are ending \emph{on} the \NS5.
In the more supersymmetric $\mathcal{N}=(2,2)$ configuration, which corresponds to the fundamental representation on all positions and no inhomogeneities in the spin chain, the \D4s are differently oriented and are located in the intervals \emph{between} two \NS5--branes (as discussed in~\cite{Orlando:2010uu}).
In this case the twisted masses of the fundamental and anti--fundamental fields are determined uniquely by the fluxtrap parameter $m$ which gets propagated to these fields via the superpotential term $Q\Phi\overline Q$, which is absent in the $\mathcal{N}=(1,1)$ case.

\bigskip

Of course, the question of how to generalize this brane construction further is of importance as the possible symmetry groups of the integrable systems captured by the Gauge/Bethe correspondence go beyond what was covered in this note. The most obvious candidates are the $B $,~$C$~and~$E$~series, but also supergroup symmetries are possible. Unfortunately, these examples are far more challenging. Quiver gauge theories are finite if and only if the quiver is an affine Dynkin diagram of type $A$,~$D$,~or~$E$~\cite{Katz:1997eq}. This spells trouble for the $B$ and $C$ series.
The approach followed here seems also not suitable to capture the exceptional groups. Studying the T--dual~\textsc{ale} singularities may be a promising alternative venue, see~\cite{Lerche:1996an}.

The generalization of the string theory construction to all symmetry groups appearing in Bethe-solvable integrable systems will thus remain a topic of further investigation for some time to come.

\subsection*{Acknowledgements}
 
It is our pleasure to thank Simeon Hellerman and Wolfgang Lerche for enlightening discussions and comments on the manuscript, and Davide Forcella for correspondence.
\renewcommand{\arraystretch}{1.2}

\begin{landscape}
  \begin{table}[t]
    \centering
    \begin{tabular}{SccSSS}
      \topline
      \tcolrow && && \multicolumn{2}{c}{\tcolcel string theory}\\
      \multicolumn{2}{c}{\multirow{-2}{*}{\tcolcel  gauge theory}} & \multicolumn{2}{c}{\multirow{-2}{*}{\tcolcel   integrable model}} & \multicolumn{1}{c}{\tcolcel \( A_r \)} & \multicolumn{1}{c}{\tcolcel \( D_r \)} \\   \midline
      number of nodes in the quiver & $r$ & $r$ & rank of the symmetry group & number of \NS5 $-1$ & number of \NS5 $+1$       \\
      gauge group at $a$-th node &$U(N_a)$ & $N_a$ & number of particles of species $a$ & \multicolumn{2}{c}{number of \D2s in the \( a \)-th stack }\\
      flavor group at node $a$& $U(L_a)$ & $L_a$ & effective length for the species $a$ & number of \D4s (\( \# \D4 \)) on the \( a \)-th \NS5  & \( 1 \le a \le r-1 \): \( L_a = \# \D4 \) in the \( a \)-th interval; \newline \( L_r = 2 \times \# \D4 \) on NO5 \( + \# \D4 \) on  \( (r-1) \)-th \NS5 \\
      lowest component of the twisted chiral superfield &$\sigma^{(a)}_i$ & $\lambda^{(a)}_i$ & rapidity & \multicolumn{2}{c}{\( (x_8 + \imath x_9) \) position of the \( i \)-th  \D2 in the \( a \)-th stack}  \\
      \colrow   twisted mass of the fundamental field & ${{}\widetilde{m}^{\fund}}_{k}^{(a)}$ & $\frac{\imath}{2} \Lambda^{a}_k + \nu^{(a)}_k$ & \( \Lambda^a_k \): highest weight of the representation  & \multicolumn{2}{T}{\colcel \( (x_6 + \imath x_7) \) separation of upper and lower part of the \( k \)-th \D4' in the \( a \)-th stack } \\
      \colrow    twisted mass of the anti--fundamental field & ${{}\widetilde{m}^{\bar \fund}}^{(a)}_{k}$ & $\frac{\imath}{2} \Lambda^{a}_k - \nu^{(a)}_k$ &  \( \nu_k^{(a)} \):  inhomogeneity & \multicolumn{2}{T}{\colcel center of mass position of the \( k \)-th \D4' in the \( a \)-th stack}   \\
      \colrow      twisted mass of the adjoint field & ${{}\widetilde{m}^{\adj}}^{(a)}$ & $\frac{\imath}{2} C^{aa}$ & diagonal element of the Cartan matrix & \multicolumn{2}{c}{\colcel fluxtrap parameter} \\
      \colrow      twisted mass of the bifundamental field & ${{}\widetilde{m}^{\bif}}^{(ab)}$ & $\frac{\imath}{2} C^{ab} $& non--diagonal element of the Cartan matrix & \multicolumn{2}{c}{\colcel\( - \frac{1}{2} \times \) fluxtrap parameter} \\
      FI--term for $U(1)$--factor of gauge group $U(N_a)$ & $\tau_a$ & $\hat\vartheta^a$ & boundary twist parameter for particle species $a$ & \multicolumn{2}{c}{\( x_3 \) separation between \NS5s in the \( a \)-th interval} \\ \bottomrule
    \end{tabular}
    \caption{%
      Dictionary for the Gauge/Bethe correspondence and the string construction for the general \( \mathcal{N}=(1,1) \) case. In the $\mathcal{N}=(2,2)$ case all the twisted masses (grey part of the table) are uniquely determined by the fluxtrap parameter $m$.%
    } 
      \label{tab:dictionary}
  \end{table}
\end{landscape}

\appendix

\section{Supersymmetry of the orbifold}
\label{sec:supersymm-orbif}

In this appendix we derive the Killing spinors preserved by the
\NO5--\NS5--fluxbrane background.

In Section~\ref{sec:d-series} we have seen that the fluxbrane
identifications are compatible with the \NO5 action. This can be
written as \( \mathcal{I}_4 \cdot (-1)^{F_L} \) where \( \mathcal{I}_4
\) is the inversion \( \mathcal{I}_4 : ( x_2, x_3, x_4, x_5 ) \mapsto
( -x_2, -x_3, -x_4, -x_5 ) \) and \( F_L \) is the left--moving
fermion number. The space is locally flat but in order to write the
orbifold action on the Killing spinors it is convenient to use polar
coordinates:
\begin{align}
  x_4 + \imath x_5 &= \rho_1 \eu^{\imath \phi_1} \, ,& x_6 + \imath x_7 &= \rho_2 \eu^{\imath \phi_2} \, , & x_2 + \imath x_3 &= \rho_3 \eu^{\imath \phi_3} \, .
\end{align}
The solutions to the type~\textsc{iib} Killing spinor equations for the
fluxbrane background in these coordinates are given
by~\cite{Hellerman:2011mv}
\begin{equation}
  K^{\text{flux}} = \proj{flux}_- \exp [ \tfrac{1}{2} \phi_1 \Gamma_{45} + \tfrac{1}{2} \phi_2 \Gamma_{67} + \tfrac{1}{2} \phi_3 \Gamma_{23} ] \epsilon_0 \equiv \epsilon_L^{\text{flux}} + \epsilon_R^{\text{flux}} \,, 
\end{equation}
where \( \epsilon_0 \) is a constant complex Weyl spinor and \( \proj{flux} \) is the projector that keeps only the spinors compatible with the fluxbrane identifications:
\begin{equation}
  \proj{flux}_\pm = \frac{1}{2} \left( \Id \pm \Gamma_{4567} \right) \, .
\end{equation}
These are the spinors that satisfy the local equations and the
fluxbrane identifications, now we have to impose the orbifold boundary
conditions.  The effect of \( \mathcal{I}_4 \) on the polar
coordinates is
\begin{equation}
  \mathcal{I}_4 : ( \rho_1, \phi_1, \rho_2, \phi_2, \rho_3, \phi_3 ) \mapsto ( \rho_1, \pi + \phi_1, \rho_2, \phi_2, \rho_3, \pi + \phi_3 ) \, .
\end{equation}
Using the fact that
\begin{equation}
  \exp [ \frac{\pi}{2} \left(\Gamma_{23} + \Gamma_{45} \right) ] = \Gamma_{2345} \, ,
\end{equation}
and adding the sign coming from the \( (-1)^{F_L} \) factor, we find that
the orbifold acts as
\begin{equation}
  \mathcal{I}_4 \cdot \left( -1 \right)^{F_L} :
  \begin{pmatrix}
    \epsilon_L^{\text{flux}} \\ \epsilon_R^{\text{flux}}
  \end{pmatrix} \mapsto
  \begin{pmatrix}
    - \Gamma_{2345} \\ &  \Gamma_{2345}
  \end{pmatrix} \begin{pmatrix}
    \epsilon_L^{\text{flux}} \\ \epsilon_R^{\text{flux}}
  \end{pmatrix} \, .
\end{equation}
At this point we need to project out the Killing spinors that are not
invariant under the action of the orbifold. This is
obtained by introducing the projectors
\begin{equation}
  \proj{NO5}_{\pm} = \tfrac{1}{2} \left( \Id \pm \Gamma_{2345} \right) \, ,
\end{equation}
which satisfy
\begin{equation}
  \Gamma_{2345} \proj{NO5}_{\pm} = \pm \proj{NO5}_{\pm} \, ,  
\end{equation}
so that the \textbf{8 orbifold--invariant Killing spinors} can be written as
\begin{equation}
  \begin{cases}
    \epsilon^{NO5}_L = \proj{NO5}_- \epsilon^{\text{flux}}_L =
    \proj{NO5}_- \proj{flux}_- \exp [ \tfrac{1}{2} \phi_1 \Gamma_{45}
    + \tfrac{1}{2} \phi_2 \Gamma_{67} + \tfrac{1}{2} \phi_3
    \Gamma_{23} ] \epsilon_0\,,  \\
    \epsilon^{NO5}_R = \proj{NO5}_+ \epsilon^{\text{flux}}_R =
    \proj{NO5}_+ \proj{flux}_- \exp [ \tfrac{1}{2} \phi_1 \Gamma_{45}
    + \tfrac{1}{2} \phi_2 \Gamma_{67} + \tfrac{1}{2} \phi_3
    \Gamma_{23} ]  \epsilon_1 \,, \\
  \end{cases}
\end{equation}
Note that \( \proj{NO5}_\pm \) are precisely the same projectors
\(\proj{NS5}_\pm \) that appears for \NS5--branes oriented in the same
directions (see Equation~\eqref{eq:susy-projectors}).

T--duality in \( x_8 \) translates this result into the \textsc{iia} orbifolded fluxtrap picture where the Killing spinors \( K^{IIA} = \epsilon_L + \epsilon_R \) are:
\begin{equation}
  \begin{cases} 
    \epsilon_L = \eu^{-\Phi/8}
    \left( \Id + \Gamma_{11} \right) \proj{NO5}_-
    \proj{flux}_- \exp [\tfrac{1}{2} \phi_1 \Gamma_{45} +
    \tfrac{1}{2} \phi_2 \Gamma_{67} + \tfrac{1}{2} \phi_3 \Gamma_{23} ] \epsilon_0 \\
    \epsilon_R = \eu^{-\Phi/8} \left( \Id - \Gamma_{11} \right)
    \Gamma_u \proj{NO5}_+ \proj{flux}_- \exp[ \tfrac{1}{2} \phi_1
    \Gamma_{45} + \tfrac{1}{2} \phi_2 \Gamma_{67} + \tfrac{1}{2}
    \phi_3 \Gamma_{23}] \epsilon_1
  \end{cases}
\end{equation}
where \( \epsilon_0 \) and \( \epsilon_1 \) are constant Majorana
spinors,
\begin{equation}
  \Gamma_u =  \frac{m \rho_1}{\Delta} \Gamma_5 - \frac{m \rho_2}{\Delta} \Gamma_7 +
  \frac{1}{\Delta} \Gamma_8 \, ,
\end{equation}
and
\begin{equation}
  \Delta^2 = 1 + m^2 \left( U  \rho_1^2 + \rho_2^2 \right) .
\end{equation}

\bibliography{ADSeries}

\end{document}